\documentstyle[aps]{revtex}

\begin{document}
\title{Addemdum to: ''The Mathematical Structure of Quantum Superspace as a
Consequence of Time Asymmetry''. }
\author{Roberto Aquilano}
\address{Instituto de F\'{\i}sica de Rosario.\\
Av. Pellegrini 250, 2000 Rosario, Argentina.}
\author{Mario Castagnino, Ernesto Eiroa}
\address{Instituto de Astronom\'{\i}a y F\'{\i}sica del Espacio.\\
Casilla de Correos 67, Sucursal 28, 1428 Buenos Aires, Argentina.}
\maketitle

\begin{abstract}
In this paper we improve the results of sec. VI of paper [M. Castagnino,
Phys. Rev. D {\bf 57}, 750 (1998)] by considering that the main source of
entropy production are the photospheres of the stars.
\end{abstract}

\section{A rough coincidence becomes more precise.}

In paper \cite{Cast} one of us reported a rough coincidence between the time
where the minimum of the entropy gap $\Delta S=S_{act}-S_{\max }$ \cite{0},
takes place and the time where all the stars will exhaust their fuel. The
time where the minimum of $\Delta S$ is located was: 
\begin{equation}
t_{cr}\approx t_0\left( \frac 23\frac{\omega _1}{T_0}\frac{t_{NR}}{t_0}%
\right) ^3  \label{14}
\end{equation}
The following numerical values were chosen: $\omega _1=T_{NR},$ the
temperature of the nuclear reactions within the stars (that was considered
as the main source of entropy), $t_{NR}=\gamma ^{-1}$ the characteristic
time of these nuclear reactions, $t_0$ the age of the universe, and $T_0,$
the cosmic micro-wave background temperature, and making some approximations
the rough coincidence was obtained.

Now we have reconsider the problem and conclude that, even if nuclear
reactions within the stars are a source of entropy, the parameters $T_{NR}$
and $t_{NR}$ are not the good ones to define the behavior of the term $%
e^{-\gamma t/2}\rho _1$ of equation (100) of paper \cite{Cast}, since they
do not correspond to the main unstable system that we must consider. In fact
the main production of entropy in a star is not located in its core, where
the temperature is almost constant (and equal to $T_{NR})$, but in the
photosphere where the star radiates. The energy radiated from the surface of
the star is produced in the interior by fusion of light nuclei into heavier
nuclei. Most stellar structures are essentially static, so the power
radiated is supplied at the same rate by these exothermic nuclear reactions
that take place near the center of the star \cite{Clayton}. We can decompose
the whole star in two branch systems\cite{0}, as explained in section VII of
paper \cite{Cast}, where a chain of branch systems was introduced. We have
two branch systems to study: the core and the photosphere. The core gives
energy to the photosphere and in turn the photosphere diffuses this energy
to the surroundings of the star, namely in the bath of microwave radiation
at temperature $T_{0\text{ .}}$In this way, we have two sources of entropy
production: the radiation of energy at the surface of the star and the
change of composition inside the star (as time passes we have more helium
and less hydrogen). Since the core of a star is near thermodynamic
equilibrium, we neglect the second and we concentrate on the first: the
radiation from the surface of the star (related with the difference between
the star and the background temperatures). So the temperature of the
photosphere and not the one of the core must be introduced in our formula.
Thus it is better to consider the photosphere as the unstable system that
defines the term $e^{-\gamma t/2}\rho _1$ of equation (100) \cite{Cast}. So
we must change $T_{NR}$ and $t_{NR}$ by $T_P$, the temperature of the
photosphere and $t_S$ the characteristic lifetime of the star. Then we must
change eq. (\ref{14}) to: 
\begin{equation}
t_{cr}\approx t_0\left( \frac 23\frac{T_P}{T_0}\frac{t_S}{t_0}\right) ^3
\label{19}
\end{equation}

As the 90\% of the stars are dwarfs with photosphere temperature $T_P=10^3K$ 
\cite{1} and the characteristic lifetime $t_S=10^9$ \cite{2} if we take
these values we reach again to.

\begin{equation}
t_{cr}\preceq 10^4t_0  \label{20}
\end{equation}
but now with no approximation. The order of magnitude of $t_{cr}$ is a
realistic one. In fact, $10^4t_0\approx 1.5\times 10^{14}years$ after the
big-bang the conventional star formation will end \cite{Adams} and it is
also considered that all the stars will exhaust their fuel \cite{AJP} so it
is reasonable that this time would be of the same order than the one where
the entropy gap stops its decreasing and begins to grow\cite{Reeves}. So the
rough coincidence it is now a precise order of magnitude coincidence and
therefore the comprehension of paper \cite{Cast} is improved.

\section{Acknowledgments.}

We wish to thank Omar Benvenutto for fruitful discussions. This work was
partially supported by grants Nos. CI1*-CT94-0004 of the European Community,
PID-0150 and PEI-0126-97 of CONICET (National Research Council of
Argentina), EX-198 of the Buenos Aires University and 12217/1 of
Fundaci\'{o}n Antorchas and the British Council.

\end{document}